\documentclass[12pt]{article}
\pdfoutput=1

\usepackage{amsmath}
\usepackage{amsfonts}
\usepackage{amscd}
\usepackage{amsthm}
\usepackage{setspace}
\usepackage{amssymb}

\usepackage{graphicx}
\usepackage{authblk}
\usepackage{caption}
\usepackage{ytableau}
\usepackage{mathtools}

\begin{document}

\begin{titlepage}

\begin{flushright}

\end{flushright}

\vskip 1cm

\begin{center}

{\bf \Large Viscosity of quark--gluon plasma and\\ gravitons Bose--Einstein condensate}

\vskip 1.2cm

Yusuke Kimura$^1$ 
\vskip 0.4cm
{\it $^1$Analytical quantum complexity RIKEN Hakubi Research Team,\\ RIKEN Center for Quantum Computing (RQC), 2-1 Hirosawa, Wako, Saitama 351-0198, Japan}
\vskip 0.4cm

\vskip 1.5cm
\abstract{We propose a theoretical model of quark--gluon plasma (QGP) produced at the Brookhaven National Laboratory (BNL) Relativistic Heavy Ion Collider (RHIC). In this model, we hypothesize that the gas of quarks and gluons are confined within the film of gravitons as a Bose--Einstein condensate (BEC) during the production of QGP. The structure of this theoretical model of QGP explains why QGP behaves in a liquid-like manner, resembling a perfect fluid rather than a gas. Based on this theoretical model, we calculated the shear viscosity of QGP. This is essentially the shear viscosity of the BEC film of gravitons. The computational results obtained in this study appear to be consistent with experimental findings.
}

\end{center}
\end{titlepage}

\tableofcontents
\section{Introduction}
Quark--gluon plasma (QGP) was produced at the Brookhaven National Laboratory (BNL) Relativistic Heavy Ion Collider (RHIC) by accelerating gold ions to ultrarelativistic speeds and colliding them \footnote{Theoretical reviews can be found, e.g. in \cite{Gyulassy2004, Shuryak2004, Jacobs2004}.}. The properties of QGP, including shear viscosity, have been experimentally analyzed. The collision melts the protons and neutrons contained within, resulting in the release of quarks and gluons that form a plasma. 

One might naturally expect that the quarks and gluons released in the collision to move freely, and the plasma to be gaseous. However, the actual property of QGP produced in the observed experiment was peculiar; it was not gaseous but rather liquid-like, and indeed QGP behaved as nearly a perfect fluid. Furthermore, its viscosity observed in the experiment is actually lower than that of any substance on Earth. 

Several considerations and investigations of QGP have been made, including efforts to compute the shear viscosity of QGP. However, the peculiarity of QGP behaving liquid-like rather than gaseous remains unexplained. 

In this work, we analyze the properties of QGP and propose a theoretical model of QGP that explains why it behaves in a liquid-like manner. Furthermore, we compute the shear viscosity of QGP based on the theoretical model constructed in this study. 

We propose a theoretical model in which the plasma of quarks and gluons is enclosed in a film of the Bose--Einstein condensate of gravitons, resulting in a liquid-like property resembling a perfect fluid. The gas of quarks and gluons captured inside the film of gravitons condensate must be gaseous, as one would naturally expect. However, because the gravitons are strongly bound together owing to the Bose--Einstein condensation \cite{Bose, Einstein1924, Einstein1925}, the plasma of quarks and gluons is unable to break free from the film of gravitons, making the film of gravitons Bose--Einstein condensate liquid-like. Due to these factors, QGP behaves close to a perfect fluid. Additionally, because a graviton does not interact with the electromagnetic field, the film of gravitons condensate surrounding the plasma of quarks and gluons went unnoticed. 

Based on this theoretical model, the shear viscosity of QGP is not due to the gas of quarks and gluons but to that of the gravitons Bose--Einstein condensate enclosing the gas of quarks and gluons. The fact that QGP exhibits a viscosity lower than any substance on Earth suggests that this low value of the viscosity is due to an unknown matter. 

In the process of protons and neutrons melting into the gas of quarks and gluons in the heavy ion collision, the binding energies constituting the major fractions of the masses of the protons and neutrons are lost. As a result of the loss of significant portions of mass in the ion collision, gravitons are released at a very close distance, at which a quantum effect becomes significant. The gravitons are expected to be released within a distance comparable to the wavelengths of the gravitons. These released gravitons, as bosons, undergo Bose--Einstein condensation, capturing the gas of quarks and gluons and forming a film. 

To demonstrate the validity of this theoretical model, we calculate the shear viscosity of the gravitons Bose--Einstein condensate (referred to as gravitons BEC) that captures the quark--gluon gas. We employ a result from \cite{Kimura2023} for this computation. While the computational result aligns with that in \cite{Kovtun2004}, the interpretation of the physics of QGP in this study differs completely from that in \cite{Kovtun2004}.The spacetime considered in this work is four-dimensional (4D), treating the elementary particles as point particles. The treatment of gravitons at the quantum level in this framework doesn't lead to a contradiction, as discussed in \cite{Kimura2023}. We believe that the proposed picture in this study provides a correct interpretation of the QGP shear viscosity. 

The properties of gravitons as matter are not well-understood. Due to this situation, computing the shear viscosity of the unfamiliar gravitons BEC requires some insight. We utilize a property of the graviton discussed in \cite{Kimura2023} for this computation. Considering a certain limit of gravitons relates to a known state, enabling us to perform a practical computation. We consider a limit in which the gravitons BEC is compressed to the Planck size. Considering this limit relates the entropy of the gravitons BEC to that of a black-hole geometry. This argument provides an estimation of the shear viscosity of QGP. The Bekenstein--Hawking formula of the black hole entropy \cite{Bekenstein1973, Hawking1975} plays a role in this computational approach. Notably, we highlight for the first time that the actual gravitons (in the context of the 4D spacetime) are relevant to a practical physical phenomenon in this study. 

This paper is structured as follows: we discuss how our theoretical model for QGP can be deduced from a physical reasoning in section \ref{sec2}. We compute the viscosity of QGP based on the theoretical model that we proposed in section \ref{sec3}. Since the gravitons BEC confines the plasma of quarks and gluons in our theoretical model, the viscosity of QGP essentially means the viscosity of the gravitons BEC. To perform the computation, we first compute the entropy of the gravitons BEC in section \ref{sec3.1}. Then, from the deduced entropy of the gravitons BEC we compute the viscosity in section \ref{sec3.2}. We state our concluding remarks in section \ref{sec4}. 

\section{Physical reasoning for quark--gluon plasma}
\label{sec2}
When heavy ions are accelerated to ultrarelativistic speeds (99.99$\%$ of the speed of light) to collide, quarks and gluons are released from the bound states that form protons and neutrons, resulting in a quark--gluon plasma. In this manner, QGP is produced in the RHIC at BNL. 

However, the produced QGP exhibited at least two peculiar properties: 
\begin{itemize}
\item[1.] QGP is fluid-like, resembling a perfect fluid.
\item[2.] QGP's viscosity is lower than that of any known substance on Earth.
\end{itemize}

The plasma of quarks and gluons, which was naturally expected to be gaseous, is observed to be fluid-like, which is odd. Furthermore, the plasma is close to a perfect fluid. This observation suggests that the plasma of quarks and gluons is enclosed in a film of something that undergoes Bose--Einstein condensation, confined within the film. 

If this film were visible, it would have been noticed. The fact that this film remains unnoticed implies that the particles constituting the film do not interact with light. Are such particles present as constituents of the heavy ions? Heavy ions do not appear to contain such a particle as a constituent. 

This means, then, that some particles arise to form the film that confines the gas of quarks and gluons during the collision process. What is significant before and after the collision is the change in the total mass. The major fractions of the masses of protons and neutrons result from the binding energy mediated by gluons that holds the constituent the quarks together. As quarks are released after the collision, the binding energy is lost, and as a consequence, the mass largely decreases. 

During the process, the large decrease in mass results in the emission of many gravitons. 

Putting these clues together, we are led to the suggestion that gravitons undergo Bose--Einstein condensation, tightly binding together to form a film that confines the gas of quarks and gluons. The quarks and gluons cannot escape from this tightly bound film of graviton condensate. This should provide a correct description of QGP. This is the theoretical model of QGP that we propose in this study.  

After the heavy ion collision, gravitons are expected to be emitted and localized within a distance which is comparable to their wavelengths, ensuring that the overlaps of the wave functions are not negligible. Quantum effect becomes important within this distance, and gravitons, as bosons, undergo Bose--Einstein condensation, forming a film that confines the gas of the released quarks and gluons. 

Bose--Einstein condensation generally occurs at very low temperatures; however, the critical temperature $T_c$ is inversely proportional to the mass $m$ of particle: $T_c\propto \frac{1}{m}$. Since a graviton is considered to be massless, gravitons undergoing Bose--Einstein condensation at a very high temperature do not present any contradiction. For gravitons, at a very high temperature, when localized at a distance that is not negligibly small compared to the overlaps of their wavelengths, it is physically reasonable to expect that gravitons undergo Bose--Einstein condensation. 

What physical consequence can one predict from this theoretical model of QGP? If nothing can be predicted, there is no way to determine whether the proposed theoretical model correctly describes QGP. Indeed, as we will demonstrate shortly, utilizing an argument relevant to the discussion in \cite{Kimura2023}, the entropy and the shear viscosity of the gravitons BEC can be computed. If the computations agree with the experimentally observed results of QGP, that means that the theoretical model of QGP proposed in this work is plausible in this respect. 

To be more concrete, we apply the proposal in \cite{Kimura2023} that a graviton gives rise to a black-hole geometry (Kerr geometry \cite{Kerr1963}, to be more precise) under certain conditions. This approach relates the entropy of the gravitons BEC to that of the black-hole geometry, and the Bekenstein--Hawking entropy formula \cite{Bekenstein1973, Hawking1975} can be applied to the entropy computation of the gravitons BEC. 

In section \ref{sec3}, we compute the shear viscosity of QGP using this approach. Our interpretation proposed in this study is that the viscosity of QGP is precisely that of the gravitons BEC surrounding the gas of quarks and gluons. The result of the computation is stated in section \ref{sec3.2}. 

\section{Viscosity of QGP and the entropy of gravitons BEC}
\label{sec3}

\subsection{Gravitons BEC and black-hole geometry}
\label{sec3.1}
Since the properties of gravitons are not well-known, at first glance, there seems to be no clue to compute their entropy and viscosity. However, considering a special limit allows us to compute these quantities. 

Here, the ``special limit'' refers to the Planck scale. We consider a situation where the gravitons BEC is contained in a region with a diameter equal to the Planck length $l_P$. In this situation, the spacing between gravitons is less than the Planck length $l_P$. Because any length shorter than the Planck length does not have a physical meaning, the spacing between gravitons in the gravitons BEC can be considered to be the Planck length $l_P$. If the wavelengths $\lambda$ of the gravitons become much less than the spacing $d$ of the gravitons ($\lambda\ll d$), the quantum aspects become less dominant in the behavior of the gravitons, and the classical description becomes effective. This implies that the gravitons come apart from the Bose--Einstein condensation. However, in the situation we just described, where the size of the gravitons BEC is within the Planck length $l_P$, the wavelength of the gravitons can be as short as the Planck length $l_P$, and the gravitons BEC still remains in the state of Bose--Einstein condensate, since the wavelength of the Planck length is comparable to the spacing of the gravitons $(d\sim l_P$, $\lambda=l_P\sim d)$. 

The discussion in \cite{Kimura2023} strongly suggests that a graviton gives rise to a black-hole geometry under a certain condition, regardless of whether it is virtual or actual. Namely, a graviton can be viewed as a source of the black-hole geometry. An actual graviton, not only a virtual graviton, can also generate a black-hole geometry under certain circumstances. Considering the Planck size region that we described previously, when gravitons in the gravitons BEC possess the wavelength of the Planck length $l_P$, it means that each graviton has an energy order of the Planck energy $E_P$. In this situation, a Planck energy is located within a region of Planck size; therefore, one expects that gravitons form a black-hole geometry. 

Because a wavelength shorter than the Planck length $l_P$ does not have a physical meaning, a graviton cannot possess a wavelength shorter than the order of the Planck length $l_P$. This implies that a graviton cannot have energy higher than the Planck energy scale. Given this, the fact that a graviton possessing energy of the Planck energy scale gives rise to a black-hole geometry is natural. This property of a graviton prohibits itself from having energy higher than the Planck scale, and the problem of the graviton's wavelength being shorter than the Planck length $l_P$ having a physical meaning can be avoided owing to this mechanism. 

Thus, for a gravitons BEC of Planck size, even when gravitons possess Planck energy, the Bose--Einstein condensation does not untangle, and the gravitons BEC in this situation is expected to generate a Kerr geometry. In this case, using the Schwarzschild radius, one can estimate the size of the formed Kerr black hole. We use the formula $r=\frac{2GE}{c^4}$ instead of $r=\frac{2GM}{c^2}$ where $E$ represents the energy of the graviton, and $r$ represents the size of a black hole. When the graviton has energy $E=E_P$, the formula
\begin{equation}
r=\frac{2GE_P}{c^4}
\end{equation}
yields 
\begin{equation}
r=2l_P;
\end{equation}
therefore, one can confirm that the size of the formed black hole is on the order of the Planck length. We conclude from these arguments that the entropy of the gravitons BEC of Planck size is well approximated by that of the black-hole geometry of the identical size. We are thus led to postulate that the entropy computed using the black-hole approximation generally yields an approximate value for the entropy of the gravitons BEC. 

We assume that the film of the gravitons BEC is a two-dimensional surface, i.e. its thickness is negligibly small compared to its size. $s$ is used to represent entropy density, and $A$ to denote the surface area of the gravitons BEC. The Bekenstein--Hawking entropy formula \cite{Bekenstein1973, Hawking1975} states that 
\begin{equation}
\label{density_bh}
s_{\rm BH}=\frac{k_B A}{4l_P^2}=\frac{c^3 k_B A}{4G\hbar},
\end{equation}
regardless of the type of black-hole geometry. What we postulated states that this well approximates the entropy density of the gravitons BEC, $s_{\rm grav.}$:
\begin{equation}
\label{bh_approx}
s_{\rm BH}\sim s_{\rm grav.},
\end{equation}
or equivalently, 
\begin{equation}
\frac{c^3 k_B A}{4G\hbar}\sim s_{\rm grav.}.
\end{equation}
We used $s_{\rm grav.}$ and $s_{\rm BH}$ to represent the entropy densities of the gravitons BEC and the black hole approximation, respectively. 

\subsection{Viscosity of QGP}
\label{sec3.2}
One finds that the relation of the shear viscosity of the gravitons BEC, $\eta$, and the absorption cross section, $\sigma(\omega)$, can be written as follows:
\begin{equation}
\eta=\frac{c^3}{16\pi G}\sigma(0).
\end{equation}

As we discussed in section \ref{sec3.1}, we utilize the black-hole-geometry approximation to compute the shear viscosity $\eta$. For computational purposes, even though the Kerr geometry is not spherically symmetric, we simplify it as if it were and proceed. Graviton's absorption cross section into a spherically symmetric black hole is identical to that of a massless scalar \cite{Kovtun2004}, therefore the result in \cite{Das1996} applies to this situation. In the black-hole-geometry approximation, according to the result in \cite{Das1996}, one has
\begin{equation}
\sigma(0)=\lim_{\omega\to 0}\sigma(\omega)=A.
\end{equation}
From these, one deduces that the shear viscosity $\eta$ can be expressed as follows:
\begin{equation}
\label{shear_visc}
\eta=\frac{c^3 A}{16\pi G}.
\end{equation}
When we calculate the ratio of shear viscosity $\eta$ \eqref{shear_visc} to entropy density $s_{\rm BH}$ \eqref{density_bh}, we obtain the following relation:
\begin{equation}
\frac{\eta}{s_{\rm BH}}=\frac{\hbar}{4\pi k_B}.
\end{equation}
Finally, utilizing the relation \eqref{bh_approx}, one deduces the following relation:
\begin{equation}
\frac{\eta}{s_{\rm grav.}}\sim \frac{\hbar}{4\pi k_B}.
\end{equation}
This gives us the ratio of shear viscosity to entropy density for QGP in our theoretical model. 

An important aspect of the theoretical QGP model proposed in this study is its ability to explain why QGP is observed to be liquid-like rather than gaseous. If QGP consisted only of quarks and gluons, it would be gaseous. The produced QGP was observed to be liquid-like because the gravitons that underwent Bose--Einstein condensation captured and confined the gas of quarks and gluons. This gravitons BEC behaves in a liquid-like manner. This theoretical model explains the aforementioned peculiar properties of QGP.

\section{Concluding remarks}
\label{sec4}
In this work, we have introduced a theoretical model of QGP, which was produced in heavy ion collisions in RHIC. According to our proposal, its true identity must be the gas of quarks and gluons confined within the film of gravitons, existing in the state of Bose--Einstein condensation. The gas of quarks and gluons, confined in the film of gravitons, cannot break free from the film. Bose--Einstein condensation typically occurs at very low temperatures; however, gravitons BEC is an exception. Since a graviton is massless, gravitons can undergo the Bose--Einstein condensation at the high temperature at which QGP is produced in the experiment. 

Based on the theoretical model we presented, we calculated the viscosity of QGP. In the model we proposed, what is observed as the viscosity of QGP is actually the viscosity of the gravitons BEC that confines the gas of quarks and gluons. Utilizing the black-hole-geometry limit, we initially computed the entropy of the gravitons BEC. From this result, we obtained the shear viscosity of QGP. The result appears to align with the experimentally observed outcome. 

According to the theoretical model proposed in this work, we have concluded that QGP behaves similarly to a perfect fluid because the gravitons BEC surrounds the plasma of quarks and gluons. 

The computed viscosity of QGP, along with the fact that the theoretical model proposed in this work explains why QGP is close to a perfect fluid, appears to support the validity of the model introduced in this work.

\section*{Acknowledgments}

Y.K. acknowledges Hakubi projects of RIKEN.

\end{document}